\documentclass[aps,pra,superscriptaddress,amsmath,amssymb,preprintnumbers,floatfix,showpacs,11pt]{revtex4}

\usepackage{amssymb}
\usepackage{epsfig}

\begin{document}
\title{ Quantum key distribution in terms of the Greenberger-Horne-Zeilinger state: multi-key generation }
\author{ Faisal A. A. El-Orany\footnote{$faisal.orany@mimos.my$\\ $Report-number$: 7.5.4}, Wahiddin M. R. B., Mustafa Afanddi Mat Nor,
Norziana Jamil, Iskandar Bahari}

 \affiliation{ Cyberspace Security Laboratory, MIMOS Berhad, Technology Park Malaysia, 57000
 Kuala Lumpur, Malaysia}

\date{\today}

\begin{abstract}
In this paper, we develop a quantum key distribution protocol
based on the Greenberger-Horne-Zeilinger states (GHZs). The
particles are exchanged among the users in blocks through two
steps. In this protocol, for three-particle GHZs three keys can be
simultaneously generated. The advantage of this is that the users
can select the most suitable key for
 communication.   The protocol can be generalized  to $N$ users
to provide  $N$ keys. The protocol has two levels for checking the
eavesdroppers. Moreover, we discuss   the  security of the
protocol against different attacks.

\end{abstract}

  \maketitle

\section{Introduction}

Quantum cryptography is one of the most fruitful applications in
the quantum information theory and it could be widely  used in the
near future \cite{Gisin}. Quantum key distribution (QKD) is one of
the main interest in the quantum cryptography, which is defined as
a procedure allowing two or many legitimate users of a
communication channel to establish two or many exact copies. This
will be in the form of one copy for each user, of a random and
secret sequence of bits. The advantage of the quantum cryptography
over the classical one lies in the following. The former follows
the quantum laws, e.g., the Heisenberg uncertainty principle,
no-cloning theorem and the quantum correlations, to protect the
distribution of the cryptographic keys. Therefor, the message and
the key are  secure since the legitimate users can easily detect
the eavesdroppers. This in turn encourages the researchers to
develop new protocols. Most of these protocols follow the original
three constructs, namely, the BB84 protocol \cite{Bennett}, the
B92 protocol \cite{Phys} and the EPR protocol \cite{Ekert}. The
QKD has been experimentally demonstrated by different means as
shown, e.g., in \cite{Breguet}.

Entanglement is one of the main ingredients in the quantum
information theory \cite{[1]}. Based on this property, various
protocols have been developed. For instance,  Ekert has introduced
his famous  protocol  which makes  two remote parties
  share a private secure key   \cite{Ekert}.
    The quantum direct communication protocol has been described
   in \cite{[19]}. The "ping-pong" protocol has been given  to achieve deterministic
direct communication between the legitimate users \cite{[20]}.
This protocol has advantages and disadvantages;  it allows the
transmission of either a secret key or the plaintext message.
Nevertheless, it is insecure (quasi-secure) if it is operating in
a noisy (perfect quantum) channel \cite{[21]}. In these protocols,
the Einstein-Polosky-Rosen state (EPR) has been
  used to distribute the quantum-cryptographic key. The security has been checked
  either by the violation of the Bell inequalities
\cite{bennt}
 or by the correlation of the EPR.
Furthermore,  Greenberger-Horne-Zeilinger states (GHZs)
\cite{Greenberger} have been already involved in the quantum
cryptography \cite{merak,buzek,gisin,zeng,Jiu,Wang,
CWang,TGao,Zhang,jwang,chamoli,comment}. These states are
distinguished by a large Hilbert space compared to  the EPR.
Precisely, in the GHZs protocols one can have two or more
legitimate users. In the two users case, the distribution of the
GHZs particles and the quantum states are asymmetrical between
Alice and Bob. Additionally, the security can be established via
the correlation of the GHZ triplet state , e.g., \cite{zeng}.
Nevertheless, in the multi-user case, say three users, there are
sender,  receiver and supervisor, who controls the entanglement
and information transmission between the sender and the receiver,
e.g., \cite{buzek,TGao,jwang}. The users can get the key only by
joint cooperation. In this respect, the protocol is secure against
the dishonest user (if he or she exists) \cite{buzek}.
 The GHZs have been used in the quantum
secure direct communication \cite{Wang,chamoli} and in the
 teleportation \cite{Jiu,TGao}.
It is worth mentioning that  the simultaneous quantum direct
comminution between users based on the GHZs has been developed in
\cite{Zhang} . Nevertheless, it has been proved that  this
protocol is not secure \cite{comment}.
  Finally, the GHZs have been
experimentally implemented by various means, e.g., using
entanglement swapping starting from three down converters
\cite{Zukowski}, using two pairs of entangled photons
\cite{Bouwmeester}, based on dipole-induced transparency in a
cavity-waveguide system \cite{junq},  in the framework of the
superconducting circuits \cite{Wei} and nuclear magnetic resonance
\cite{Nelson}.

In this communication, we develop a new protocol using the GHZs.
More illustratively, we have three or more legitimate users and no
controllers. The number of the  keys, which can be generated in
this protocol, equals the number of the users and/or the number of
particles  in the GHZs. In the proposed  protocol, we apply the
block-data transfer among the users  \cite{Wang}. There are some
basic differences between this protocol and the others that were
given earlier \cite{buzek,gisin,zeng,Jiu,Wang,
CWang,TGao,Zhang,jwang,chamoli}, as we will show below. Firstly,
the legitimate users can simultaneously generate various keys.
Each user does not need the cooperation of the other users to
obtain these keys. Eavesdroppers can be checked in the two stages
of the protocol, which makes the protocol highly secure. It is
worth mentioning that  the simultaneous quantum direct commination
between users based on GHZs has been established in \cite{Zhang}.
In this protocol, the users can obtain the messages only when
Ailce announces publicly the forms of the initial and final
states. Based on this  announcement, Eve can directly obtain most
of the messages without using potential attacks. It is enough for
her to compare the initial and final states \cite{comment}. This
type of attack is called information leakage attack. The protocol,
we present in this paper, is secure against such type of  attack,
as we will show next. Moreover, we discuss   the security of the
protocol against different types of attacks.

Before  starting the description of the protocol we shed some
light on the set of the three-particle GHZs. This set includes
eight independent states, which form a complete orthonormal basis
as follows:

\begin{eqnarray}
\begin{array}{lr}
  |\psi_1\rangle=\frac{1}{\sqrt{2}} (|000\rangle+|111\rangle),\quad
|\psi_2\rangle=\frac{1}{\sqrt{2}} (|000\rangle-|111\rangle),
   \\
   |\psi_3\rangle=\frac{1}{\sqrt{2}} (|100\rangle+|011\rangle),\quad
|\psi_4\rangle=\frac{1}{\sqrt{2}} (|100\rangle-|011\rangle),
\\
|\psi_5\rangle=\frac{1}{\sqrt{2}} (|010\rangle+|101\rangle),\quad
|\psi_6\rangle=\frac{1}{\sqrt{2}} (|010\rangle-|101\rangle),
\\
|\psi_7\rangle=\frac{1}{\sqrt{2}} (|110\rangle+|001\rangle),\quad
|\psi_8\rangle=\frac{1}{\sqrt{2}} (|110\rangle-|001\rangle).
\label{1}
\end{array}
\end{eqnarray}
These states can be switched into each other by applying one of
the four unitary operators $\hat{I}^{(j)},\hat{\sigma}_x^{(j)},
i\hat{\sigma}_y^{(j)}, \hat{\sigma}_z^{(j)}$ to them, where the
superscript $j$ stands for the $j$th-particle and the notations
have the same standard meaning in the literatures. This fact is
the main object in generating the keys. The users have to agree,
in advance, about the following Boolean values:

\begin{equation}
\hat{I}^{(j)}\longrightarrow 0,\quad
 \hat{\sigma}_x^{(j)}\longrightarrow 1.
\label{2}
\end{equation}
There is one fact we would like to address here: why we do not
consider the four-encoding processes
$\hat{I}^{(j)},\hat{\sigma}_x^{(j)}, i\hat{\sigma}_y^{(j)},
\hat{\sigma}_z^{(j)}$? Actually, sometimes  applying two
operations on  the states $|\psi_j\rangle$ can give the same
results as follows:
\begin{eqnarray}
\begin{array}{lr}
\hat{I}^{(2)}\hat{I}^{(3)}|\psi_1\rangle=\hat{\sigma}_z^{(2)}\hat{\sigma}_z^{(3)}|\psi_1\rangle=|\psi_1\rangle\\
\\
\hat{\sigma}_x^{(2)}\hat{\sigma}_x^{(3)}|\psi_1\rangle=\hat{\sigma}_y^{(2)}\hat{\sigma}_y^{(3)}|\psi_1\rangle=|\psi_3\rangle
\label{1an}
\end{array}
\end{eqnarray}
This, of course, could be a weak point in the decoding process,
however, it is a positive point in  the  security of the protocol
since it  confuses eavesdroppers.

 Now we are in a position to
describe the protocol, which will be discussed in greater details
for three users, namely, Alice, Bob and Charlie. The protocol goes
as follows.

 Step 1 : Alice  prepares   a sequence  $A$ of $n + d+d'$ ordered GHZ
triplets, each of which has the form $(a_1^k,a_2^k,a_3^k), \quad
k=1,2,...,n+d+d'$, where the superscript $k$ denotes the order of
the triplet in the sequence $A$. These triplets are randomly
chosen from the set given by (1), and  are already known to Alice
herself. Alice takes one  particle $a_1,a_2,a_3$ from each GHZ
triplet to form three ordered particles sequences
$A_j=\{a_j^1,a_j^2,...,a_j^{n+d+d'}\}$ with $j=1,2,3$.

Step 2: At the same time Bob and Charlie, in their sites, do the
same as Alice. In this case Bob and Charlie have the sequences
$B_j=\{b_j^1,b_j^2,...,b_j^{n+d+d'}\}$ and
$C_j=\{c_j^1,c_j^2,...,c_j^{n+d+d'}\}$ with $j=1,2,3$,
respectively.

Step 3: Alice transmits  the sequences $A_2$ and $A_3$ to Bob and
Charlie, respectively. Similarly, Bob transmits the sequences
$B_1$ and $B_3$ to Alice and Charlie, respectively.  Charlie
transmits $C_1$ and $C_2$ to Alice and Bob. Each sender  should
inform, via classical channels,  the receivers  before the
transmission and the receivers should confirm  the reception of
the particles. This process is used to avoid  unwanted
circumstances under which Eve can impersonate one or both of the
users. The transmission of the particles occurs in blocks and the
orders of the sequences are not known for the receivers. These
arrangements may increase the security of the protocol.

Step 4: The users start to check the security of the channels to
see if the eavesdroppers are on  line or not. This should be
independently  performed for each sequence. We start with the
Alice's sequence $A=(A_1,A_2,A_3)$. Bob, say, chooses randomly a
large subset $d$ of particles from the sequence $A_2$ and measures
each of them using one of the two  bases $x$ or $z$. Then Bob
publicly tells Alice and Charlie via a classical channel about the
positions, the basis and the measurement outcome for each of the
particles. Charlie, using the same bases, measures the
corresponding particles from the sequence $A_3$, and publicly
tells the others about the results. After that Alice applies the
same procedure for the  particles in the home sequence $A_1$.
Consequently, the users can decide whether there are eavesdroppers
in the line or not. Precisely, if the measurement outcomes are
different, then Eve is not on  line.
 Similar procedures have to be executed for the sequences $\{B_1,B_3\}$ and $\{C_1,C_2\}$.
At this step, there is no need  to evaluate the error rates since
the keys have not been generated yet. The final remark we should
stress here is that the user, who has the home particles, should
be the last one  executing eavesdropper checking for the set $d$.
This, in turn, helps in finding out the dishonest user if he or
she exists.

Step 5: At this stage, we assume that each partner has his own
key, and he or she wants to transmit it  to the others. Therefore,
each one encodes his own key in the particles of the other
partners by means of the operations shown in (\ref{2}).  Each one
uses one operator to act simultaneously on the two particles from
the different sequences, however, in the same order. For instance,
suppose that Alice wants to encode the bit $1$ in the
$j$th-particle for the other partners. In this case, she should
act by $\hat{\sigma}_x^{(j)}$ on the $j$-th particle in the
sequences $B_1$ and $C_1$. Similar procedures have to be performed
in Bob and Charlie sites. During the encoding process the users
should be careful regarding the information and the positions of
the particles of the set $d'$. The   reason is that the users will
sacrifice these particles when checking the eavesdroppers in the
final step.

 Step 6: After completing the encoding process each user transmits back
the blocks of particles (message particles) to the other partners.
Of course they should inform each other before transmission and
after the reception of the blocks. At this stage the users
 obtain their original particles, but in  new forms. For
instance, Alice, after a successful transmission, has the sequence
$A=\{(a_1^1,a'_2,a'_3),(a_1^2,a_2^{'2},a_3^{'2}),...,(a_1^{n+d'},
a_2^{'n+d'},a_3^{'n+d'})\}$, where the dash means that these
particles are different from the original ones since now they
carry the keys. It is worth mentioning that the sequences
$\{a_2^{'j},j=1,...,n+d'\}$ and $\{a_3^{'j},j=1,...,n+d'\}$
include  Bob and Charlie keys, respectively. However,
$\{a_1^{j},j=1,...,n+d'\}$ is the home sequence. As Alice, say,
prepared these particles initially she knows them very well. Now
Alice performs the GHZs-basis measurement on the ordered $n+d'$
GHZs and compares the measurement outputs with the initial forms
of the states to obtain the keys of Bob and Charlie. For instance,
if Alice initially prepared one of the triplet in the state
$|\psi_1\rangle$ and  the measurement result is $|\psi_3\rangle$.
From (1) and (2) Alice has the relation
$|\psi_3\rangle=\hat{\sigma}_x^{(2)}\hat{\sigma}_x^{(3)}|\psi_1\rangle$.
According to the arrangement (2) Alice knows with certainty that
Bob and Charlie bits are $1$ and $1$, respectively, and so on. At
the same time Bob and Charlie perform the same procedures. At the
end of the protocol each user has three keys: his own key and the
keys of the other partners. At this moment there is no overlap
between the users since each one has retrieved  his own sequence
of the GHZs, however, in the new forms. Thus they cannot check the
eavesdroppers based on the entanglement property, e.g.,
\cite{bennt}. In this case, the users can use the virtue  of the
set  $d'$.
 As we have mentioned above, the senders know the positions of the particles
 of the set $d'$ and the information
included. Precisely, they designed their keys based on that these
bits  will be excluded from the generic keys. To explain this
scenario we focus the attention on the set $d'$ of Alice. Alice
publicly informs Bob and Charlie via a classical channel about the
set $d'$, i.e. the positions of the particles and the bases which
should be used for the measurement but not the results. Bob and
Charlie follow Alice's prescription and they announce the
measurement results sequentially. In other words, Bob announces
the measurement of the first particle, then Charlie the second
one, then Bob and so on. Such process is sufficient to detect  the
dishonest user if any. We proceed, if there are overlaps between
the results of the measurements then the key is secure otherwise
they should evaluate the error rate for this key. Then the users
should follow the same steps for the sequences $B$ and $C$.
Comparison among these three error rates drives the users to
choose the most suitable  key for the communication. Moreover,
based on the positions of the particles in the set $d'$ the users
can bring some diversions  to delude Eve, i.e. by using particular
types of swapping entanglement and/or shifting process for the
bits of the final keys. Nevertheless, they should agree in advance
about this scenario.

This protocol can be extended to  $N$ parties. In this case, at
the end of the protocol, the users obtain $N$  keys. Each user
generates a sequence of $N$ particles in the GHZs, which has the
form:
\begin{equation}\label{1n}
 |\psi\rangle=\frac{1}{\sqrt{2}}(|j_1 j_2...j_N\rangle+|j'_1
 j'_2...j'_N\rangle),
\end{equation}
where $j_p,j'_p, (p=1,...,N)$ are bits $0$ or $1$ according to the
specified states. The users should follow the same steps discussed
above. In this protocol, the users  obtain the generic key by
making a comparison among the $N$ keys. The larger the numbers of
the keys, the higher the probability of obtaining a secure key.
This is related to the fact that Eve cannot efficiently attack
many keys at the same time. She  is vulnerable   enough at least
for one of the keys.

Now we comment on the security of the proposed protocol. From the
above discussion, it is obvious that we have two levels of
security; before and after the encoding process. Generally, Eve
can get information on the keys if she manages to obtain
information about the particles before and after the encoding
process.  This, of course, requires that the users did not detect
her in the  step $4$. Suppose that we deal with ideal conditions
and   Eve managed to attack the travelling particles. We start the
discussion with the double-CNOT attack. This attack does not
disturb the channels and hence the mutual information between
different users are unity, i.e., $I_{ij}=1$ where $i,j=A,B,C,E$.
This means that if the users are going to use only a message
authentication as a security strategy, they would never be able to
detect this kind of attack. The mechanism of the 2CNOT attack is
as follows: In the forward path, Eve performs a first CNOT gate
between the particles in transit from Alice to Bob and
 to Charlie (control qubits) and her ancillae (target qubits).
The second CNOT gate is executed in the backward path. Restricting
ourselves to Alice's particles, the scenario of the 2CNOT attack
can be explained as follows. Alice keeps the first particle in her
site  and sends the second and third ones to Bob and Charlie,
respectively. Eve executes her ancillae
$|0\rangle_2^{E}|0\rangle_3^{E}$ with the transit particles and
performs the first CNOT gate as:
\begin{equation}\label{9}
U_{cont}(|\psi_1\rangle|0\rangle_2^{E}|0\rangle_3^{E})=|\Psi_1\rangle=
\frac{1}{\sqrt{2}} (|0\rangle |0\rangle |0\rangle_2^{E}
|0\rangle|0\rangle_3^{E}+|1\rangle |1\rangle |1\rangle_2^{E}
|1\rangle|1\rangle_3^{E}).
\end{equation}
It is obvious that the CNOT gate creates an entangled state
composed from the travelling qubits and the Eve's ancillae.
Suppose that Bob and Charlie act, respectively, by $\hat{I}^{(2)}$
and $\hat{\sigma}_x^{(3)}$ on their corresponding  particles
according to their own keys. Thus we obtain:
\begin{equation}\label{10}
\hat{I}^{(2)} \hat{\sigma}_x^{(3)}|\Psi_1\rangle=|\Psi_2\rangle=
\frac{1}{\sqrt{2}} (|0\rangle |0\rangle |0\rangle_2^{E}
|1\rangle|0\rangle_3^{E}+|1\rangle |1\rangle |1\rangle_2^{E}
|0\rangle|1\rangle_3^{E}).
\end{equation}
In the backward path Eve executes the second CNOT gate, which
leads to:
\begin{equation}\label{11}
U_{cont}(|\Psi_2\rangle)=|\Psi\rangle= \frac{1}{\sqrt{2}}
\left(|0\rangle |0\rangle
 |1\rangle+|1\rangle |1\rangle |0\rangle\right)|0\rangle_2^{E}
 |1\rangle_3^{E}=|\psi_7\rangle|0\rangle_2^{E}|1\rangle_3^{E}.
\end{equation}
It is evident that the generic state is flipped, i.e.
$|\psi_1\rangle\longrightarrow|\psi_7\rangle$.  In addition, the
generic state and the Eve ancillae are disentangled. For Eve, it
is enough to measure these ancillae in the $z$ basis to get the
information. It is obvious that  this information, in some sense,
could be sufficient to give Eve information regarding the keys.
 We should stress that Eve cannot know the forms of the generic
backward states since she cannot access the home particles.  If
Eve knows the Boolean relations given by (\ref{2}), then she can
obtain the keys. This is quite similar to the non-entangled
protocol discussed in \cite{marco} for which the 2CONT attack is
reasonably efficient. Generally, Eve does not know any information
about the agreement (\ref{2}) and hence she should take into
account the actions of the operators $\hat{\sigma}_z^{(j)}$ and
$i\hat{\sigma}_y^{(j)}$.  In this regard, the probability to get
the key from this protocol, via the 2CNOT attack, is $25\%$. This
is clear since the applied operation could be, e.g., one of the
set
$\{\hat{I}^{(2)}\hat{I}^{(3)},\hat{I}^{(2)}\hat{\sigma}_z^{(3)},
\hat{\sigma}_z^{(2)}\hat{I}^{(3)},
\hat{\sigma}_z^{(2)}\hat{\sigma}_z^{(3)}\}$. Accordingly, this
indicates that the 2CONT attack cannot give Eve valuable
information about the keys.

 We draw the attention to the quantum man-in-the-middle
attack. The mechanism of this attack can be explained for Alice's
particles as follows. In the forward path, i.e, Alice transmits
the particles to Bob and Charlie, Eve blocks the the particles,
stores them in the memory and sends instead her particles to Bob
and Charlie. In the backward path, i.e. Bob and Charlie transmit
the particles back to Alice, Eve measures these particles to get
the encoded information. At this moment she encodes the same
information in the stored particles  and passes them back to
Alice. This is a dangerous attack, in particular, when Eve has
full control of the classical channel. The solution against this
kind of attack is that the legitimate users should share a prior
secret letting them authenticate the channel and make it reliable
when they communicate before and after the transmission processes.
For the protocol under consideration, this attack is not helpful
for Eve, where the users can detect her through the forward path
in the step 4 with high probability. For instance, suppose that
Alice uses the state $|\psi_1\rangle$, where she keeps the first
particle with her and sends the second and third particles to Bob
and Charlie, respectively. Eve has two possibilities for choosing
ancillae, namely, GHZs and $z$ basis states.  If Eve uses one of
the GHZs as a faked state, according to the set (1), Eve has a
probability of $1/8$ to use the correct states. In this case, she
has a probability of $1/3$ to keep the correct particle with her
pretending to be Alice.  In this case, the probability that Eve is
not revealed at all in a step 4, is $1/24$. On the other hand, if
Eve is not on the line, the measurements of the users in a control
run (for the state $|\psi_1\rangle$) yield $|000\rangle$ and
$|111\rangle$ with equal probability \cite{Zhang}. Now suppose
that Eve uses the faked states from the $z$ basis states, so what
would be the probability to be detected in a control run?. If  Eve
qubits are one pair of the set
$\{|0\rangle_B|0\rangle_C,|1\rangle_B|1\rangle_C\}$, where the
subscripts $B$ and $C$ indicate that these particles are sent to
Bob and to Charlie, respectively, the measurement results
corresponding to these  ancillae  are
$\{(|000\rangle,|100\rangle),(|011\rangle,|111\rangle)\}$. Thus
Eve will be detected, because her eavesdropping introduces an
error rate equal to $1/2$. When  Eve's  ancillae are one pair of
the set $\{|0\rangle_B|1\rangle_C,|1\rangle_B|0\rangle_C\}$ the
users detect her with certainty. This shows that the protocol is
secure against this attack.

Information leakage attack has been developed in \cite{comment}
and can be explained as follows: In some of the GHZ protocols,
e.g. \cite{Zhang}, the users obtain the keys only  when Alice
announces publicly its initial and final states. Thus  Eve can
easily obtain these results and  when comparing them she obtains
the keys. Actually, it has been proved, based on this attack, that
Eve can obtain three bits from the transmitted four bits
\cite{comment}. It is obvious that the protocol under
consideration does not include such type of announcements and
hence it  is secure against this type of attack.
 We conclude by referring to the  intercept-resend and disturbance
 attacks. They have been already studied in \cite{Zhang} for the GHZ protocol
 and showed that it is secure against these attacks. Similar
 arguments can be quoted for the proposed protocol.
Finally, the security of  the ping-pong protocol against
considerable quantum channel losses is discussed in \cite{[21]}.
The current protocol can be treated in a quite similar way. This
is based on the fact that it is enough   for Eve to attack one of
the travelling sequences of each users, e.g. $A_2, B_1,C_3$.
 If the  attention is focused  on one qubit of each sequence, the
 treatment will be the same as that of the ping-pong protocol \cite{[21]}.

In conclusion, we have presented a protocol based on the
entanglement property of the GHZs. After a complete round, the
users would have many keys so that they can choose the relevant
one. There is no announcements about the forms of the states used
in the protocol. This  improves the efficiency of the QKD.  The
proposed protocol is a two-way: In the forward process the senders
transmit blocks of particles to the receivers, who encode the keys
in these particles and transmit them back (backward process) to
the senders. Eavesdroppers have been checked through two  stages.
 The security  has been
discussed  against different attacks. An important remark, which
guarantees the security of this protocol against considerable
number of attacks, is the entanglement. For protocols operating
via entanglement, the secret information is encoded in the whole
entangled state. Consequently, Eve cannot get useful information
if she has obtained just a part of the entangled state. In the
under consideration protocol  Eve cannot access the particles of
the home sequences, i.e., the particles of
 $A_1,B_2,C_3$.

%%%%%%%%%%%%%%%%%%%%%%%%%%%%%%%%%%%%%%%%%%%%%%%%%%%%%%%%%%%%%%%%%
\section*{References}
%%%%%%%%%%%%%%%%%%%%%%%%%%%%%%%%%%%%%%%%%%%%%%%%%%%%%%%%%%%%%%%%%

\end{document}